\documentclass[useAMS,usenatbib]{mn2e}
\pdfoutput=1
\usepackage{ifthen}
\RequirePackage{amsmath}
\RequirePackage{amssymb}
\usepackage{graphicx}

\newcommand{\bfe}{\mbox{\bf e}}

\newcommand{\bfJ}{\mbox{\bf J}}
\newcommand{\bfB}{\mbox{\bf B}}

\newcommand{\ephi}{\mbox{$\bfe_\phi$}}

\newcommand{\Dif}[2]{\frac{\mbox{$\partial #1$}}{\mbox{$\partial #2$}}}

\def\bfnabla{\mbox{\boldmath $\nabla$}}

\def\bfOmega{\mbox{\boldmath $\Omega$}}

\newcommand{\Curl}{\bfnabla \times}

\def\eps{\mbox{$\varepsilon$}}

\newcommand{\bt}{\begin{tabular}{@{}p{2cm}p{14.5cm}}}

\def\references{\section*{References}
    \bgroup\parindent=1cm\parskip=1mm\small
    \def\refpar{\par\hangindent=1em\hangafter=1}\reference\ \vskip-10mm}
\def\endreferences{\refpar\egroup}

\def\reference{\relax\refpar \ \hskip-1.1cm} 

\newboolean{draft}
\newcommand\figlab[1]{\ifthenelse
{\boolean{draft}}
{\;\;\fbox{\bf{fig:#1}}}{}\label{fig:#1}}
\newcommand\mylabel[1]{\ifthenelse
{\boolean{draft}}
{\;\;\fbox{\bf{eq:#1}}}{}\label{eq:#1}}
\newcommand{\Eq}[1]{Equation~(\ref{eq:#1})}
\newcommand{\Fig}[1]{Figure~\ref{fig:#1}}
\newcommand{\bfig}{\begin{center}\begin{figure}}
\newcommand{\efig}{\end{figure}\end{center}}
\newcommand{\FIG}[4]{\bfig\centerline{\includegraphics[width=#2cm]{#1}}
  \caption{#3}\label{fig:#4}\efig}

\newcommand{\FIGFOUR}[7]{\bfig\centerline{\includegraphics[height=#5cm]{#1}\qquad\includegraphics[height=#5cm]{#2}}\vskip  -2 cm \centerline{\includegraphics[height=#5cm]{#3}\qquad\includegraphics[height=#5cm]{#4}}
  \caption{#6}\label{fig:#7}\efig}
\newcommand\mytitletext{\ifthenelse{\boolean{draft}}{ -- {\bf DRAFT}}{}} 
\newcommand\mydatetext{\ifthenelse
{\boolean{draft}}
{\bf Draft Date -- \clock{\time} on \today}{Under consideration for MNRAS}}
\ifthenelse{\boolean{draft}}{\parskip =28 truept}{}

\newcommand{\average}[1]{\ensuremath{\langle #1 \rangle}}

\newcount\h 
\newcount\m 
\newcount\w
\def\clock#1{
\w=#1
\m=\w
\divide \w by60 
\h=\w
\multiply \w by60
\advance\m by-\w 
\ifnum\h<10 0\fi\number\h\thinspace :\thinspace\ifnum\m<10 0\fi\number\m}

\title[Effects of fluctuation on  $\alpha\Omega$ dynamo models]{Effects of fluctuation on $\alpha\Omega$ dynamo models}
\author[Michael R.E.Proctor]{Michael R.E.Proctor\thanks{E-mail:
mrep@cam.ac.uk;} \\
Centre for Mathematical Sciences, University of Cambridge,Wilberforce Road, Cambridge CB3 0WA, UK}
\begin{document}

\date{Accepted 2007 August 22. Received 2007 August 21; in original form 2007 July 19.}

\pagerange{\pageref{firstpage}--\pageref{lastpage}} \pubyear{2007}

\maketitle

\label{firstpage}

\begin{abstract}
We analyse the role of a fluctuating $\alpha$-effect  in $\alpha\Omega$ dynamo models, and show that there is a mechanism for magnetic field generation, valid at large scale separation, deriving from the interaction of mean shear and a fluctuating $\alpha$-effect. It is shown that this effect can act as a dynamo even in the absence of a mean $\alpha$-effect, and that the timescale for dynamo waves is strongly affected by the presence of fluctuations.
\end{abstract}

\begin{keywords}
helicity; mean field dynamo
\end{keywords}

\section{Introduction}
The mean-field {\it ansatz} has for many years been an invaluable tool in the construction of tractable dynamo models for the Sun, stars and planets. Extended treatments of the theory may be found in the monographs of \citet{b4} and \citet{b3}. In its traditional form, the theory distinguishes sharply between helical and non-helical turbulence. The latter case is presumed to imply that the statistics of the small-scale velocity field that ultimately produces the mean-field effects are symmetric under reflection; this leads to the vanishing of the $\alpha$-effect term in the mean induction equation. Any mean inductive effects of a more general sort (for example, the "$\bfOmega\times\bfJ$" effect of R\"adler) are associated (like the diffusion term) with two spatial derivatives of the magnetic field, and are not, unlike the $\alpha$-effect, guaranteed to lead to mean field growth on sufficiently large scales.

Investigation of the dynamo properties of velocity fields which are non-helical {\it on the average}, however, has usually ignored the possible effect of fluctuation. It is well known that even non-rotating turbulence has large fluctuations of helicity, though the mean is essentially zero. Even if the sign of the helicity is well defined, there is no guarantee that the emf's generated from a mean field by the helical flow will not have large fluctuations. {The paper of \cite{b9}, which made a systematic attempt to derive transport coefficients from a simulation of shear flow turbulence, noted the presence of significant fluctuations.} This effect is even more graphically illustrated by the results of a recent numerical experiment on convection in a rotating layer \citep{b1}. A weak uniform horizontal field is imposed on a fully developed convective flow in the layer, and the emf measured. It is found that although the helicity, when averaged over the upper part of the convective layer, has a definite sign, the emf fluctuates wildly in magnitude and direction and if there is a non-zero average it is very small compared with the rms fluctuations. \Fig{fig0}, from their paper, shows this clearly.

\FIG{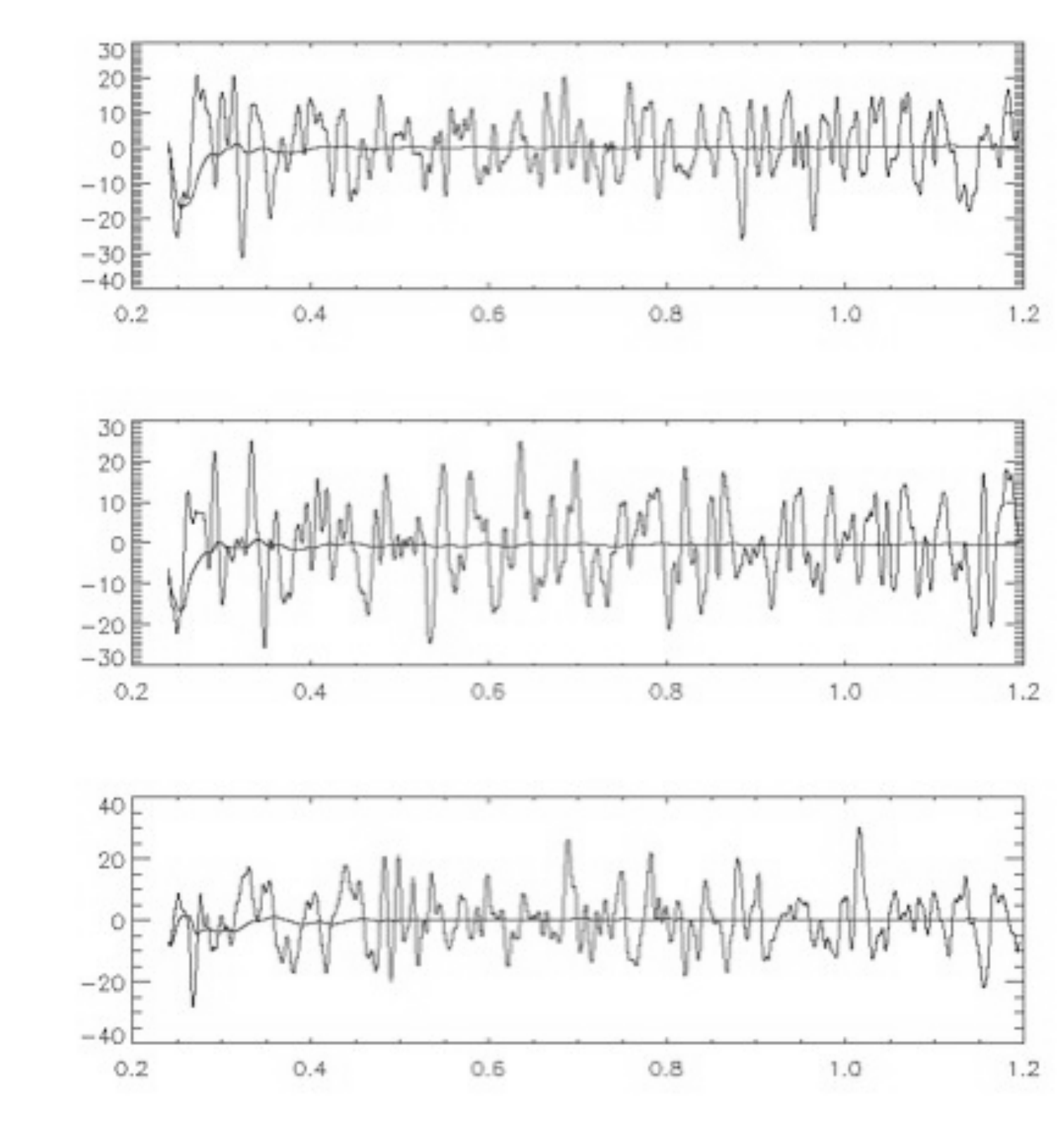}{8}{Time series showing the behaviour of the mean emf (averaged in the horizontal and over the top half of the layer) for the rotating convection simulation of Cattaneo \& Hughes 2006). The three plots show the measured emf in each of the three coordinate directions ($z$ is vertical). The heavier line shows the running time-average. The imposed magnetic field is in the $x$ direction.}{fig0}

It is of interest, therefore, to understand whether large fluctuations in the induced mean emf, on a timescale slower than the original averaging process but faster than the timescale for mean field evolution, can lead to significant dynamo action. A number of authors have considered this problem from various points of view. An early investigation was made some time ago by \citet{b2} (see also \citet{b4}, Ch.~7), who examined an $\alpha^2$ dynamo, with no mean flow, in which the $\alpha$-effect exhibited fluctuations with zero mean. The analysis, based on first order smoothing, led to a possible negative diffusion effect, giving rise to the possibility of mean field growth. In the case of negative diffusion, however, the 
consequent rapid growth of small scales would violate the scale separation on which the averaging process was based. Thus although very interesting, the results could not be said to be entirely conclusive.

The Cattaneo and Hughes calculation does not possess large scale coherent shear, and the situation is therefore somewhat like the scenario of Kraichnan, which started from the supposition of an $\alpha^2$-dynamo model. Indeed Cattaneo and Hughes found no large-scale dynamo mode in their calculation, in spite of the large rms $\alpha$.  The situation appears rather different when there is a coherent zonal shear, such as occurs in the Sun. In that case there is a powerful mechanism for creating zonal from meridional field, and this might make the effects of fluctuating helicity more important.

There have been a number of attempts to address this situation. \cite{b6} investigated a reduced mean field model appropriate to galactic dynamos. This takes the form of an ODE and the fluctuating $\alpha$ effect plays the role of multiplicative noise. They show numerically that growth can occur for large enough fluctuations. They also report on a numerical experiment on a spatially extended model, but do not attempt a systematic survey. These ideas were further investigated by \cite{b8} from the point of view of stochastic calculus. An approach close to that of the present paper was adopted by \cite{b7}. After very lengthy and detailed analysis it was concluded that the effect of fluctuations was a term of $\alpha$-effect type, derived from spatial variations of the fluctuations.

The purpose of this paper is to revisit the problem using a simpler formulation than that of Silant'ev . We show that the situation is not quite as he envisaged. There is in fact a mechanism that leads to  dynamo action, even when the mean value of $\alpha$ vanishes, and the fluctuations have no spatial dependence. The mechanism is guaranteed to work for sufficiently large-scale mean fields, in contrast to Kraichnan's mechanism. In view of the  numerical demonstrations described above, which show that relying on a well defined $\alpha$ may be misleading, it would seem that this new mechanism may well be of great importance in understanding the dynamo properties of turbulent flows.

We demonstrate the mechanism by starting with the standard $\alpha\Omega$ dynamo equations, and allow $\alpha$ to vary rapidly in time. A simple asymptotic theory yields the new {\it ansatz}. The results are then applied to a simplified one-dimensional dynamo model to show how dynamo action is enhanced by the new mechanism. Finally we give results from a nonlinear one dimensional dynamo simulation incorporating the new mechanism, and demonstrate the important effect that the fluctuations have on the frequency of activity cycles. The paper concludes with suggestions for future development.

\section{Derivation of the mean field equations}

In order to explain the mechanism we reduce the model to its simplest form. Consider an axisymmetric mean field $\bfB=B(r,\theta)\ephi+\bfB_p[\equiv\Curl (A(r,\theta)\ephi)]$ in spherical polar coordinates $(r,\theta,\phi)$. The only mean flow is that of zonal shear with differential rotation $\Omega(r,\theta)$. Due to small scale fluctuations  there is a zonal emf proportional to $B$, of the form $\alpha(r,\theta,t)B$. {Here and throughout the fluid is taken to have uniform magnetic diffusivity $\eta$}. The induction equation then takes the standard form for an $\alpha\Omega$ dynamo,
\begin{align}
\Dif{A}{t}&=\alpha B+\eta D^2 A, \label{eq:aeq}\\
\Dif{B}{t}&=r\sin\theta\bfB_p\cdot\nabla\Omega+\eta D^2B,\label{eq:beq}
\end{align}
where $D^2=\nabla^2-1/r^2\sin^2\theta$. It is now envisaged that $\alpha$ varies in time on a timescale $\tau$ that is rapid compared with the evolution time of the mean field (this can always be achieved provided that the scale of the field is large enough). We then define a small parameter $\eps$ and write
\begin{equation}
\Dif{}{t}\rightarrow \Dif{}{t}+\eps^{-1}\Dif{}{\tau};~~\alpha=\alpha_0+\eps^{-1}\alpha_1(\tau), ~{\rm with}~ \average{\alpha_1}=0,
\end{equation}
and $\average{\cdot}$ denotes an average over the fast timescale. We also define $B=B_0+\eps B_1(\tau)$, $A=A_0+ A_1(\tau)$, where the dependence on $r,\theta,t$ has been suppressed. The leading order fluctuating parts of  Equations (\ref{eq:aeq},\ref{eq:beq}) then give
\begin{equation}
\Dif{A_1}{\tau}=\alpha_1 B_0,~~~\Dif{B_1}{\tau}=r\sin\theta\bfB_{1p}\cdot\nabla\Omega.
\end{equation}
When this is solved for $A_1,B_1$ we can calculate the order 1 quantity $\average{\alpha_1 B_1}$ that will appear in the average equation for $A_0$. This is conveniently given in terms of the function $\gamma(\tau)$, defined by $\partial\gamma/\partial\tau=\alpha_1$, $\average{\gamma}=0$; we find
$\bfB_{1p}=\Curl{\gamma B_0 \ephi}$ and
\begin{equation}\begin{split}
\average{\alpha_1 B_1}&=-\average{\gamma B_{1\tau}}=-\average{\gamma r\sin\theta\Curl{\gamma B_0\ephi}}\cdot\nabla\Omega\\
&=-\average{\gamma^2}\nabla(r\sin\theta B_0)\times\ephi\cdot\nabla\Omega\\&\qquad-\frac{1}{2}\nabla\average{\gamma^2}\times r\sin\theta B_0\ephi\cdot\nabla\Omega
\end{split}\label{eq:avterm}\end{equation}
The mean field equation for $A$ can now be obtained from \Eq{aeq} by averaging, to yield
\begin{equation}
\Dif{A_0}{t}=\average{\alpha_1B_1}+\alpha_0 B_0+\eta D^2 A_0, \mylabel{aveq}
\end{equation}
where the averaged term is given by \Eq{avterm}. The mean version of \Eq{beq} is unaltered except for subscripts zero on $\bfB_p,B$.

We can explain this effect physically as follows: Consider a zonal magnetic field varying linearly in latitude. Then the action of a spatially uniform but time dependent $\alpha$-effect is to produce a spatially uniform radial field that also oscillates. This is turn is acted upon by the zonal shear to produce an oscillating zonal field. This is acted on by the fluctuating $\alpha$-effect and the resulting zonal emf has a non-zero time average.

It is not difficult to incorporate the effects of spatial variation of the fluctuations into teh analysis. The results are not qualitatively affected by this generalization, provided that the effects of spatial diffusion are smaller on average than that of the time dependence. 
   
It will be shown in the next section that the new terms in \Eq{aveq} are capable of producing growing fields even when $\alpha_0=0$. We can see immediately that both the new mean term in \Eq{aveq} and the flux stretching term in \Eq{beq} have only one space derivative in $B_0$, $A_0$ respectively. This means that at sufficiently long length scale these terms will dominate over the diffusion terms.

\section{A one-dimensional dynamo wave}

Rather than conduct a fully two-dimensional numerical experiment, we can show how the new term leads to field growth by considering a one dimensional wave model of the type originally proposed by Parker (see, for example, \citet{b5}). Specifically, we consider $A$ and $B$ to be functions of $x,t$ alone, and consider the model
\begin{align}
\Dif{A}{t}&=\alpha B+\eta (A_{xx}-\ell^2 A), \label{eq:asim}\\
\Dif{B}{t}&=\Omega' A_x+\eta (B_{xx}-\ell^2 B),\label{eq:bsim}
\end{align}
Here $\alpha$ and $\Omega'$ are independent of $x$, though $\alpha$ is a function of time in the manner discussed above. Applying exactly analogous analysis to this equation, we obtain the following simplified version of \Eq{aveq} (the subscripts zero on $A,B$ have now been dropped)
\begin{equation}
\Dif{A}{t}=-G^2\Omega' B_x+\alpha_0 B+\eta (B_{xx}-\ell^2 B), \mylabel{avsim}
\end{equation}
where $G^2=\average{\gamma^2}$. This equation can now be solved together with \Eq{bsim}, to find marginal solutions of the form $A,B\propto \exp i(kx+\omega t)$. Substituting into equations (\ref{eq:bsim},\ref{eq:avsim}) yields the dispersion relation
\begin{equation}
\left(i\omega+\eta(k^2+\ell^2)\right)^2=G^2\Omega'^2k^2+i\Omega'\alpha_0 k. \mylabel{disp}
\end{equation}
The imaginary part $\omega=\Omega'\alpha_0 k/2\eta (k^2+\ell^2)$ (so that the dynamo waves travel if $\Omega'\alpha_0\ne 0$), while the real part can be written in dimensionless form. If we write $k=\ell m$, and define
\begin{equation}
\mathcal{D}=\frac{\Omega'\alpha_0}{\eta^2\ell^3},~~~\mathcal{Q}^2=\frac{G^2\Omega'^2}{\eta^2\ell^2},
\end{equation}
then $\omega=\eta\ell^2\mathcal{D}\cdot m/2(1+m^2)$ and $\mathcal D$, $\mathcal Q$ and $m$ are related by
\begin{equation}
\mathcal{Q}^2m^2+\mathcal{D}^2\frac{m^2}{4(1+m^2)^2}=(1+m^2)^2.
\label{eq:marg}
\end{equation}
When $\mathcal Q=0$ we have the usual $\alpha\Omega$ dynamo. Dynamo action is possible when $\mathcal{|D|}\geq 32/3\sqrt{3}$, with equality when $m=1/\sqrt{3}$. Even when $\mathcal{D}=0$, we can find dynamo action when $\mathcal{|Q|}\geq 1/\sqrt{2}$, with equality when $m=1$. Although in this case $\omega=0$ so that the wave does not travel, the example shows that the fluctuations in $\alpha$ alone can lead to growing magnetic fields, even when there is no mean emf. The stability boundary is shown in \Fig{fig1}.

\FIG{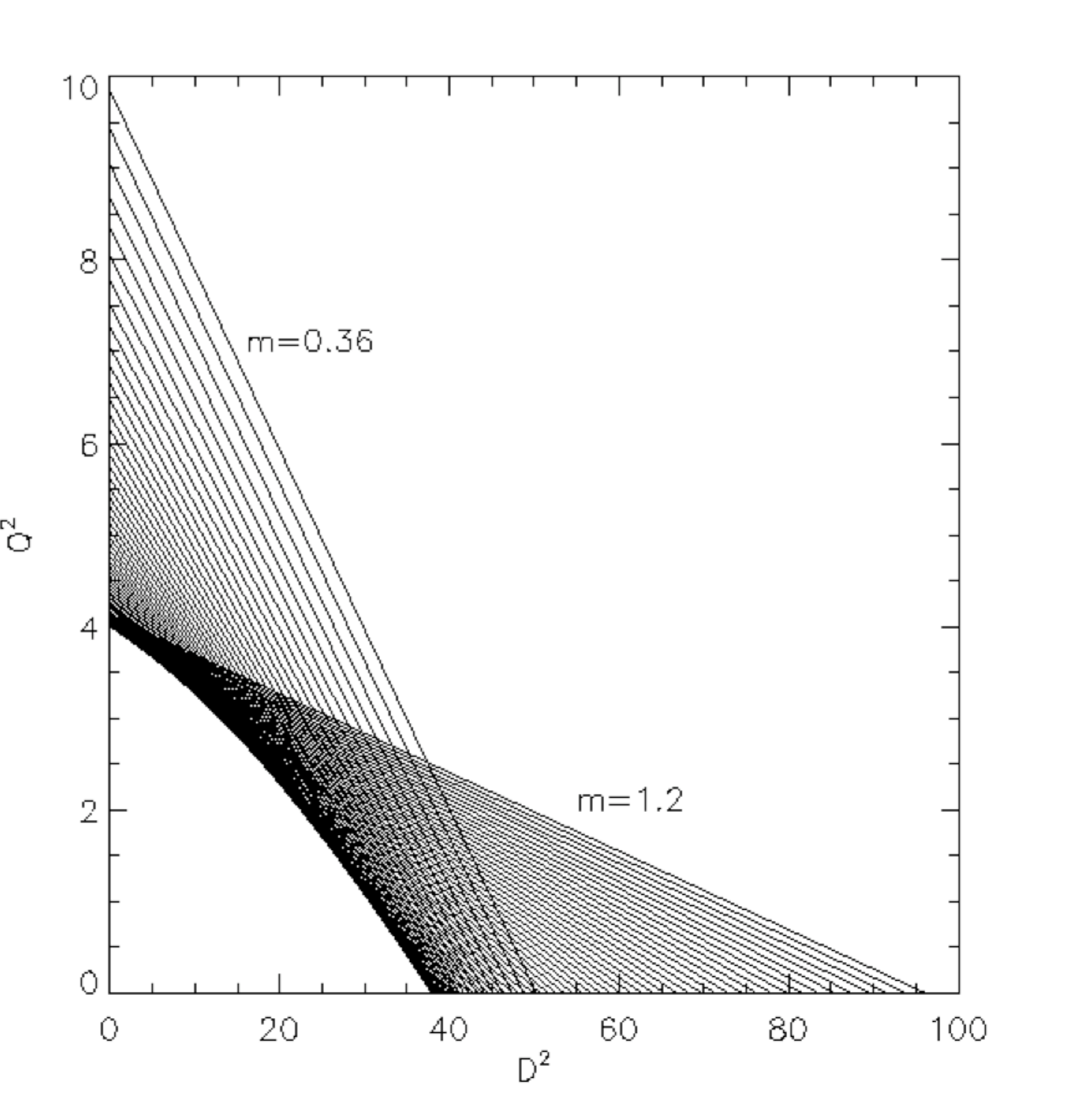}{8}{Relation between $\mathcal{D}^2$ and $\mathcal{Q}^2$ for marginal stability according to \Eq{marg}. Each straight line is for a constant value of $m$, between $m=0.36$ and $m=1.2$. Dynamo action is possible above the envelope of the lines.}{fig1}

\section{A simple model of the solar activity cycle}
For dynamical reasons it has long been accepted that any $\alpha$-effect should be antisymmetric about the solar equator. {The new term is related to the mean squared value of $\alpha$ (actually to the mean square of its time integral), and thus on average (the statistics of $\alpha$ being odd, and so those of $\alpha^2$ even about the equator), one should take it as symmetric about the equator.} This then, as one might hope and expect, preserves all the parities of conventional models. To investigate cyclical behaviour we solve numerically a model consisting of \Eq{bsim}  with $\Omega'=\eta=\ell=1$ together with the following version of \Eq{avsim}:
\begin{equation}
\Dif{A}{t}=-\frac{rB_x+d\sin (2\pi x/l) B}{1+B^2}+ B_{xx}- B), \label{eq:newavsim}
\end{equation}
where $d,r$ are  positive constants. We have adopted the simplest possible quenching formula so as to ensure equilibration of the mean fields. These equations are solved in $0<x<l$, where $A,B$ are made to vanish at $x=0,l$. It is easy to find time-periodic solutions of dipole parity with $B$ antisymmetric about $x=l/2$ . In \Fig{fig2} are shown space-time plots of $B(x,t)$ for $l=10$, $d=30$ and various values of $r$. It can be seen that as $r$ increases and fluctuations become more important the period of the cycle rises rapidly, and eventually the solution becomes steady.

\FIGFOUR{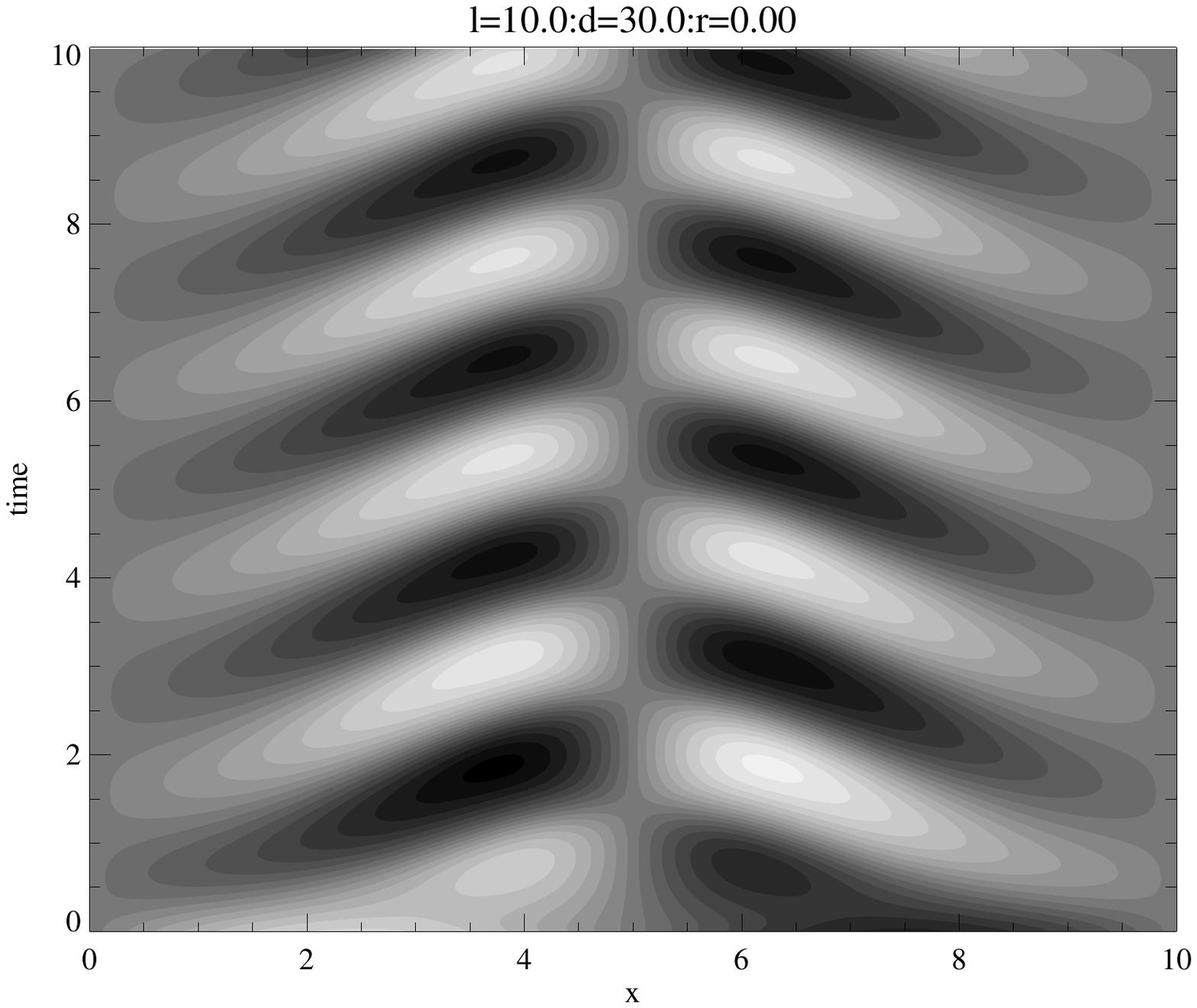}{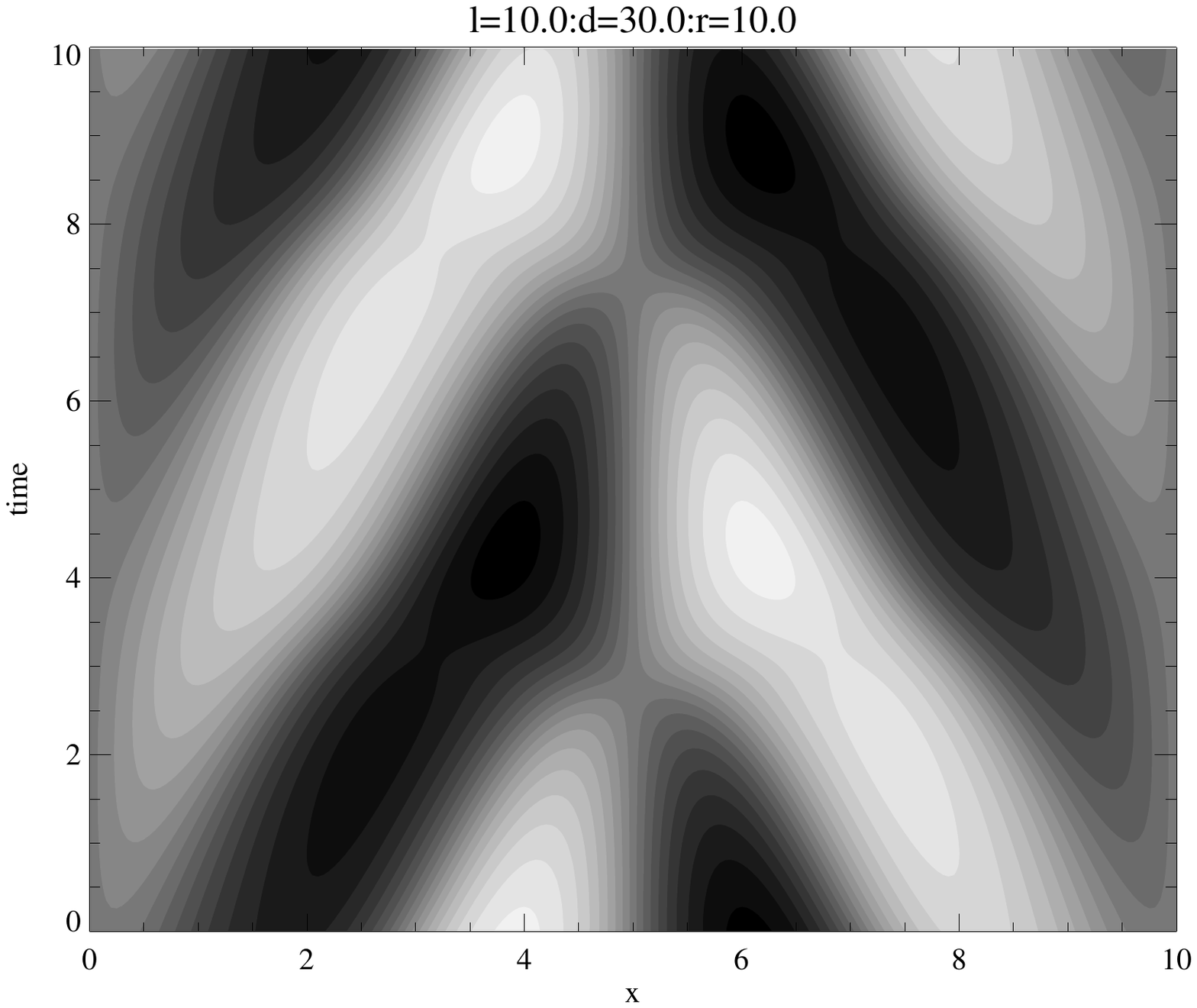}{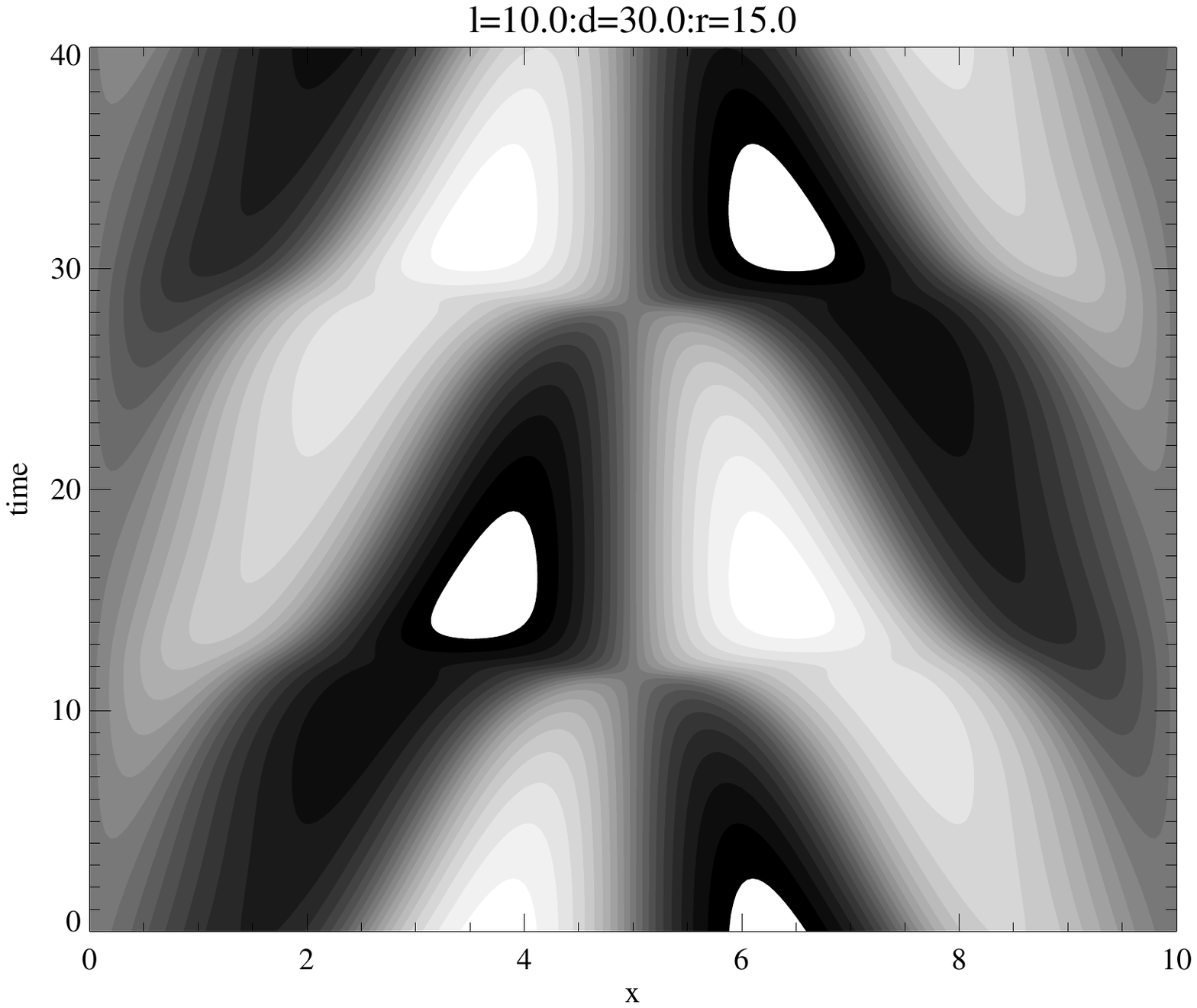}{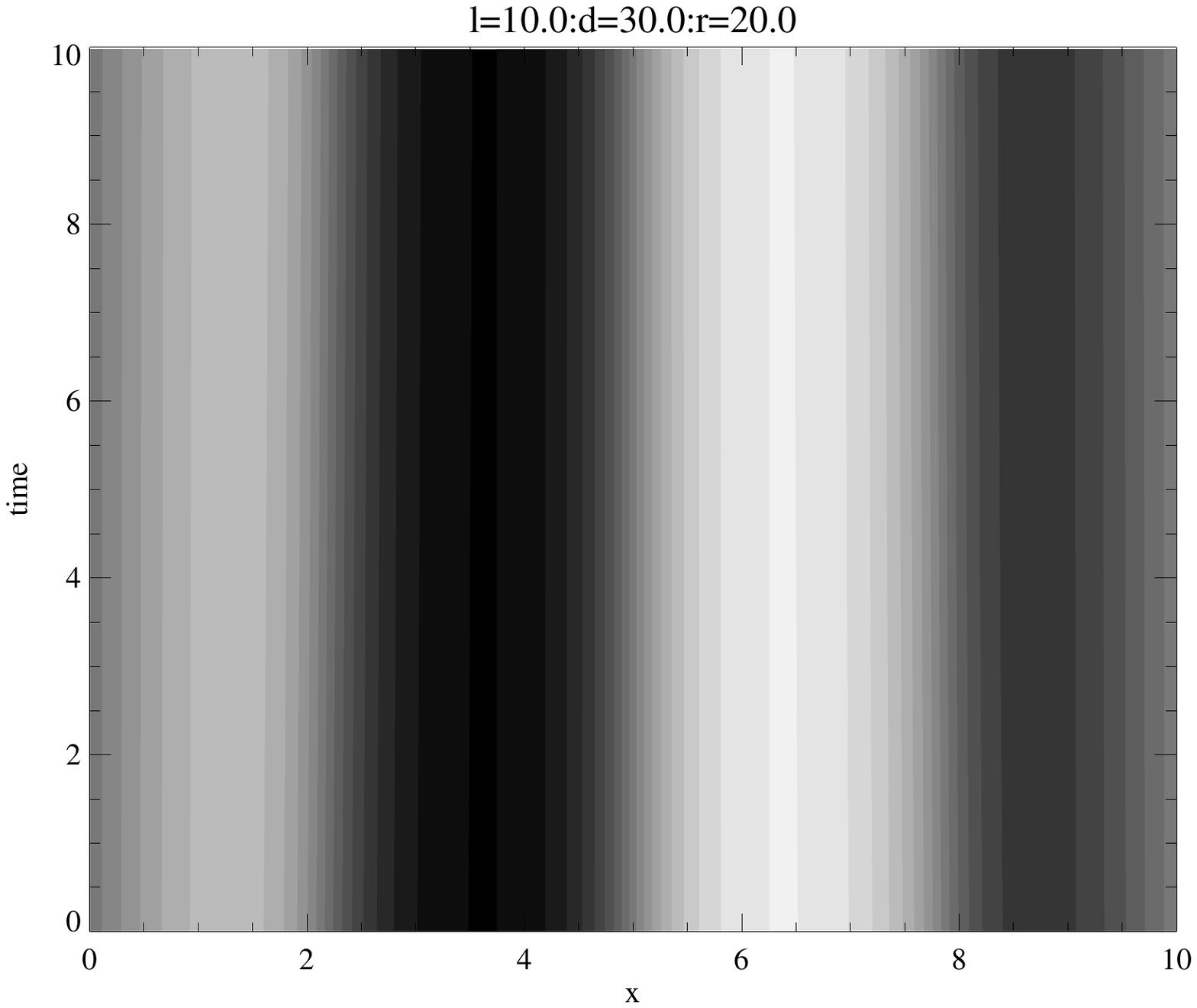}{5}{Space time plots of $B(x,t)$ for solutions of  Equations (\ref{eq:bsim}) ,(\ref{eq:newavsim}) for $l=10$, $d=30$ and four different values of $r$. For $r=0$ we have the usual $\alpha-\Omega$ dynamo. The cycle time increases with $r$ (note the different timescale in the third image)  for $r$ large enough the solution becomes steady as shown in the final picture.}{fig2}

\section{Discussion}
In this short paper it is shown that the interaction of shear with fluctuating mean field effects leads to a dynamo mechanism that may well prove more powerful than the usual $\alpha$-effect in astrophysically interesting situations, though the cyclical behaviour of any model will continue to depend on the usual $\alpha$-$\Omega$ interaction. While the derivation of the mean field equations has necessarily involved particular scalings for the amplitudes of the fluctuations, it should be emphasized that these are in some sense arbitrary as the mechanism can be guaranteed to give dynamo action at long enough scales, however small the effect. It is now necessary to integrate more realistic models in spherical geometry, and to derive the results for the mean field directly from the velocity fields; these, and the effects of spatial fluctuation, are the subject of work in progress.

\section*{Acknowledgments}

I am grateful to H.K.Moffatt and D.W.Hughes for helpful discussions.

\bsp

\label{lastpage}

\end{document}